# Beyond deficit-based models of learners' cognition: Interpreting engineering students' difficulties with sense-making in terms of fine-grained epistemological and conceptual dynamics

Ayush Gupta[a] & Andy Elby[a,b]
Departments of [a]Physics and [b]Curriculum and Instruction,
University of Maryland, College Park, MD 20742.
(The authors contributed equally to this work)

## Abstract

Researchers have argued against deficit-based explanations of students' troubles with mathematical sense-making, pointing instead to factors such as epistemology: students' beliefs about knowledge and learning can hinder them from activating and integrating productive knowledge they have. In this case study of an engineering major solving problems (about content from his introductory physics course) during a clinical interview, we show that "Jim" has all the mathematical and conceptual knowledge he would need to solve a hydrostatic pressure problem that we posed to him. But he reaches and sticks with an incorrect answer that violates common sense. We argue that his lack of mathematical sense-making—specifically, translating and reconciling between mathematical and everyday/common-sense reasoning—stems in part from his epistemological views, i.e., his views about the nature of knowledge and learning. He regards mathematical equations as much more trustworthy than everyday reasoning, and he does not view mathematical equations as expressing meaning that tractably connects to common sense. For these reasons, he does not view reconciling between common sense and mathematical formalism as either necessary or plausible to accomplish. We, however, avoid a potential "deficit trap"—substituting an epistemological deficit for a concepts/skills deficit—by incorporating multiple, context-dependent epistemological stances into Jim's cognitive dynamics. We argue that Jim's epistemological stance contains productive seeds that instructors could build upon to support Jim's mathematical sense-making: He does see common-sense as connected to formalism (though not always tractably so) and in some circumstances this connection is both salient and valued.

## Introduction

Engineering students often have troubles with mathematical problem solving in their courses. Many instructors and commentators attribute these difficulties to a deficit of skills and/or knowledge: the student is assumed to need training and practice in certain mathematical and problem-solving skills or learn the proper concept [1-3]. Others argue that many students who have difficulty with mathematical problem solving possess the needed skills and knowledge, but fail to activate that knowledge in appropriate combinations because of their epistemologies — their views about the nature of knowledge and learning [4],[5]. For example, a student who thinks that mathematical equations and conceptual ideas are two disjoint kinds of knowledge that don't "speak" to each other might fail to activate his conceptual knowledge when thinking about how to use a given mathematical equation [6].

A great deal of research supports the view that math and science students' epistemologies play a major role in explaining their difficulties learning and using mathematics effectively [7-11]. This paper, a case study of a student who gets stuck while solving a problem during a clinical interview, adds to that literature in three ways—all three of which connect to the goal of more nuanced descriptions of "what impedes learners' ability to learn with understanding" [12].

First, in addition to showing that "Jim's" difficulties stem in part from his epistemology, we also show that skill/knowledge deficits play *no* role in his difficulties. This point warrants attention from engineering education researchers and practitioners because, at first glance, his difficulty appears to stem from exactly the kind of knowledge/skills deficit commonly used to explain difficulties such as Jim's.

Second, our case study provides a fine-grained explanation of *how* Jim's epistemology affects his mathematical problem-solving in a particular episode. Illustrating a specific mechanism of this influence complements the large-*N* studies showing that certain coarse-grained epistemological beliefs correlate with certain approaches to learning and using mathematics.

Third, unlike previous literature that describes students' epistemologies in terms of hard-and-fast beliefs or stages [6],[7],[13], we show that Jim's problematic epistemological stance is nuanced and can undergo subtle shifts in response to contextual cues. This point has instructional implications: instead of needing to confront and replace problematic epistemological views such as the one Jim initially manifests, instructors can create instructional environment that tend to trigger and stabilize the more productive aspects of the epistemological views that student already has.

## Connection between Epistemologies and mathematical sense-making

In this section we define mathematical sense-making for the purposes of this paper and then briefly review previous work exploring how certain epistemological views can support or hinder students' mathematical sense-making.

### Mathematical sense-making

According to engineers, engineering educators, and engineering education researchers the effective use of mathematics in engineering cannot be reduced to a list of formal manipulations and problem-solving skills [9],[14-20]. A solid skill set is crucial, of course; but it must be integrated into a productive approach toward learning and using mathematics in engineering, an approach involving *mathematical sense-making* [9],[14-16],[21],[22]. Broadly defined, mathematical sense-making involves looking for meaning and coherence (i) within the mathematical formalism itself and (ii) between the math and the system it describes [6],[11],[21-23]. This paper focuses on (ii): the crux of mathematical sense-making, as we will use the term, is the propensity and ability to translate back and forth between mathematical relations on the page and causal or functional relations in world. By contrast, in their physics classes, many science and engineering majors treat equations as mere problem solving tools to memorize, associated with algorithms to be practiced [11],[24].

### Links between epistemologies and mathematical sense-making

Intuitively, we might conjecture that mathematical sense-making rests on the epistemological belief that equations and operations *can* be made sense of, that equations "say" something [6],[11],[22]. We now discuss previous work that, taken as a whole, supports this conjecture by showing that students' epistemological views affect whether they engage in mathematical sense-making.

One line of research shows connections between students' epistemological views and their learning of math [9],[11],[26],[25]. For example, Schommer, Cruse and Rhodes [25] administered a survey of students' epistemologies and a survey of their study strategies. They also had students read a passage about statistics. Students then took a test of their comprehension of that material. Schommer et al. [25] found that a strong correlation between students' comprehension of the passage and their epistemological views about whether knowledge is simple (piecemeal) vs. complex (interconnected). The effect of epistemology on learning remained even when the authors controlled for previous knowledge of and exposure to mathematics (using SAT scores and previous courses taken) and gender. Furthermore, the students' epistemological views correlated with their self-reported study strategies. Although this study does not probe students' mathematical sense-making *during problem solving*, it strongly suggests that epistemology affects whether students take a sense-making approach to learning mathematical ideas.

Schoenfeld [10],[11],[27], who did much of the early work in this area, also documents a connection between certain belief abut math and rote approaches to learning and problem solving in math. For example, Schoenfeld [27] investigated college students' approaches to geometry via clinical interviews in which students were presented with questions on geometric proof and geometric construction. Schoenfeld found that students relied on guesses rather than deductive ideas to solve a construction problem even though they could correctly demonstrate the deductive proofs when asked. Schoenfeld relates this behavior to a belief that "the processes of formal mathematics (e.g. proof) have nothing to do with discovery or invention [10]. Schoenfeld [10] also analyzed high school classroom activities to support his argument that curricula and assessments, even in top-rated schools, encourage and reward the epistemological stance that learning math is a matter of memorizing disconnected algorithms. For example, preparing students (successfully!) for the New York Regents exam in math, in a unit on geometric constructions, the teacher emphasized quick, rote production of standard constructions. There was no discussion of how those constructions connect to the proofs students had been doing just weeks before.

Other threads of Schoenfeld's work [28] establish the importance of metacognition in sophisticated mathematical problem solving. Students at the beginning of his mathematical problem solving class tended to choose a problem-solving path and get stuck there, even when the path was unproductive. By contrast, mathematicians — and Schoenfeld's students after a semester in his class — tended to monitor whether a given pathway was productive, and they spent more time deliberating with themselves or with other students about the pros and cons of various approaches, not just at the beginning of a problem but throughout the process. We mention this work because whether a student uses his metacognitive abilities connects closely to his epistemological stance. For instance, if a student doesn't think equations are supposed to make sense, he's unlikely to ask himself whether an equation is making sense to him.

Complementing large-N research such as Schommer et al.'s [25] and Schoenfeld's [27], case studies delve more deeply into individual students' reasoning, elucidating causal mechanisms by which epistemology affects mathematical sense-making during problem solving.

One study, of students in an introductory calculus-based physics course at Berkeley, documented how some students consider it appropriate to suspend their common sense about the physical world when engaged in mathematical problem solving [6]. All six of the interviewed subjects were comfortable with and adept at mathematical manipulations, *and* they were able to reason conceptually. However, four students displayed the epistemological view that physics knowledge (for non-experts, at least) consists mostly of disconnected pieces and that concepts are merely cues for the "real" knowledge of physics, which is equations. Those four students tended to use math differently from the other two, who viewed physics knowledge as conceptual (but expressible in equations) and as coherently interconnected.

A particularly clean illustration of the difference comes from comparing Roger and Tony, who initially made the same mistake when solving for the acceleration of block 1 in the frictionless modified Atwood machine shown here (Figure 1).

Both Roger and Tony found the force along the direction of motion to be $F = m_1 g - m_2 g \sin(\theta)$. Then they both proceeded, incorrectly, to separately connect that force to each block separately, by writing $F = m_1 a_1$ and $F = m_2 a_2$. Their calculations then yielded different accelerations for the two blocks, a result both students considered counterintuitive. At this point, the two students' reasoning diverged. Roger, who espoused and displayed a "physics is piecemeal formulas" stance throughout the six hours of interviews, checked the algebra and decided he was "90% sure" that his calculations were correct. By contrast, Tony, who through six hours of interviews espoused and enacted a view of physics as conceptual and coherent, stuck with his physical intuition that the blocks must have the same acceleration. Finding no mistake in his algebra, he dug deeper found the error in the equations he had written, realizing that his *F* refers to the net force on the two-block system, not the force on either block alone.

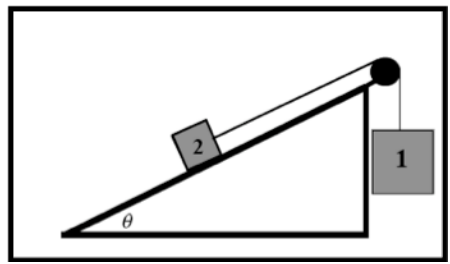

This episode was indicative of their general approach towards using mathematics. Tony strived to achieve congruence between mathematics and common sense reasoning while Roger, though considering such congruence as desirable, chose to reject common sense when it conflicted with formal knowledge. We should note that, throughout multiple problem-solving episodes, Tony and Roger were equally adept at mathematical manipulations. The sophistication in Tony's problem-solving behavior, i.e., his deep engagement in mathematical sense-making, stems in part from his epistemological view that mathematical equations express meaning that can be reconciled with common sense. Unfortunately, Roger's approach to mathematical problem solving was typical among students at Berkeley [6],[29]. Redish [24] had similar experiences with students at Maryland.

Another case study of a student's difficulty with mathematical sense-making comes from Lising and Elby [30], who like Hammer conducted six hours of interviews with a given student but who also observed the student's classroom behavior during physics tutorials (collaborative active learning sessions). They argued that "Jan's" difficulty understanding some concepts and equations stemmed not from a lack of conceptual or mathematical ability but from "a component of her epistemology, her perception of a 'wall' between formal reasoning and everyday/intuitive reasoning." [30] In one particular classroom discussion, while working with three other students on a tutorial designed to help students reach a conceptual understanding of the electric field equation $E = F/q$, Jan repeatedly comes to erroneous conclusions. For example, at one point, she claims that the electric field depends on the test charge. At another moment, she says the field remains the same with increasing distance from the source charge. With each, mistake, her group mates not only correct the particular error but (re)explain the meaning of $E = F/q$, where $q$ is a "test charge." Each time, Jan says she understands, and she seems to believe it; at other points, she readily admits when she doesn't understand something. So why does she have trouble here? Not because she just plugs and chugs; in her reasoning, she attends to the proportionalities and inverse proportionalities between the different variables. And not because of inability or unwillingness to reason conceptually; in interviews, she repeatedly displays good conceptual reasoning. What she fails to do, in class and in interviews, is connect her mathematical understanding to physical cause-and-effect relationships. So, she continues to ignore the fact that increasing $q$ also increases $F$, even when corrected four times by her group mates. Lising and Elby explain Jan's troubles with $E = F/q$ partly in terms of her robust epistemological stance that formal and everyday/intuitive knowledge don't "speak" to each other.

Jim, the student in our case, also has epistemological views that can impede him from integrating his everyday/intuitive and mathematical reasoning, as we'll argue below.

**Epistemological variability in the context of mathematical problem solving**

In explaining students' mathematical behavior, the research discussed above attributes a unitary epistemology: the relevant belief or stance is shown or assumed to be robust and consistent, at least within the contexts probed by the study. Jim, however, exhibits behavior that is best explained (we argue) by attributing epistemological variability: different contextual cues and mental states trigger different aspects of his epistemology to become foregrounded. We now briefly review the literature on epistemological variability in math and science students.

Redish, Hammer and colleagues [31-34] have been formulating and demonstrating the explanatory power of a cognitive theory that accounts for epistemological variability in terms of the context-dependent activation of locally coherent networks of epistemological "resources" (pieces of knowledge about knowledge). Building on this foundation, Bing and Redish [26] developed a framework for analyzing how upper division physics majors justify their reasoning while solving mathematical physics problems. They divided the kinds of warrants (justifications) used by students into four major categories:

(1) *Invoking authority*, often by quoting a rule such as "work is path independent";

(2) *mathematical consistency*, where a regularity in the mathematical formalism is used to justify a move or conclusion, e.g., two opposite charges behave like two nearby asteroids because both systems follow inverse square laws;

(3) *calculation*, where an answer is justified by its emerging from a trustworthy algorithm, correctly applied; and

(4) *physical mapping*, which focuses on the connections and coherence between the mathematical formalism and the described physical system.

Category (4) corresponds to the aspect of mathematical sense-making we're focusing upon in this paper, while (2) corresponds to another aspect of mathematical sense-making. In Bing & Redish's framework, these four classes of warrants reflect four epistemological stances. For our purposes, two key empirical findings from Bing & Redish deserve emphasis. One, almost every student studied showed epistemological variability, switching amongst the different epistemological stances indicated by the four classes of warrants. Two, students often got "stuck" for too long within a given epistemological stance and a corresponding approach to the problem, rather than asking themselves if another approach might be more productive. Taken together, these two findings show that what looks in the moment like a stubbornly-held epistemological belief about mathematics can turn out to be a *locally* stable epistemological stance, with the student capable of switching to another epistemological stance in response to external or internal conceptual or social cues. We will argue that Jim's seemingly stubborn epistemological stance (when he gets stuck) can shift in response to conceptual cues.

**Gaps in the engineering education literature**

Except for some of Hammer's [6] interview subjects, all of the research discussed above focuses on math and science students, not engineering majors. Within the engineering education literature, we found few articles that directly and empirically address how students use mathematics in engineering contexts. This existing research focuses mostly on characterizing broad patterns in the student behaviors or their attitudes towards mathematics.

Atman et al. [35] explore how college freshmen and seniors approach open-ended design problems. One result was that both groups of students spend very little time evaluating their results and designing alternative solutions. In this way, they resembled Schoenfeld's [28] "beginner" students who used limited meta-cognition. Evaluating results and attending to alternate strategies connect to students' sense of what "using mathematics" *is*. Atman et al. do not explore in depth what led the students not to allocate time or effort to evaluation and alternative solutions.

Like us, Fadali et. al. [17] explore whether students' attitudes towards mathematics deserve more blame for students' difficulties than mathematics deficiencies do. So far, their efforts have been directed towards design and validation of an attitudes assessment tool.

More recently, Cardella [18],[36] has explored engineering undergraduate and graduate students' mathematical reasoning in Capstone design projects and as part of long-term projects in industry partnerships. Cardella has emphasized the need to incorporate students' beliefs and attitudes along with the mathematical knowledge base to better understand students' activities in these contexts.  But this work does not explore the fine-grained dynamics of context and cognition that guides students' moment-to-moment decisions.

Though many studies acknowledge the need to improve how students use mathematics in problem-solving, engineering design, and modeling [37-39], there are few engineering education articles that analyze the details of students' reasoning during problem solving (or when they are stuck) in real time [18].  Our case study helps to fill this gap, in a way that complements the work of Atman et al.[35], Cardella[18],[36], and Fadali et al.,[17] by fleshing out one causal mechanism by which a student's epistemological stance can hinder mathematical sense-making.

## Methods
Our qualitative research methods reflect both the goals of this study and the limitations of all case studies.  First, our analysis cannot significantly add to the evidence *that* epistemologies affect how students learn and use math, but it can illustrate one mechanism for *how* the effect occurs.  This illustration can add to previous arguments [6],[25],[30] for increasing the role of epistemology in our diagnoses of student difficulties with learning to use mathematics in science/engineering problems.  Second, this work illustrates (but cannot prove, as a general rule) the explanatory power of viewing epistemologies not as globally robust beliefs or theories that consistently drive students' learning and problem solving, but rather, as context-dependent local coherences of thought whose stability depends on external cues and on the student's conceptual and even emotional state.

### Background information and subject selection
Jim is one of seven engineering majors we clinically interviewed in fall 2008, for the NSF-funded project *Improving students' mathematical sense-making in engineering:  research and development*.  All were taking the freshman first-semester calculus-based introductory physics course at a large mid-Atlantic State university.  The interviewer was not involved in teaching the course.  In all cases, we saw evidence of epistemology affecting the student's use of mathematics.  Jim's was the first interview we transcribed and analyzed in detail.  We selected him for this case study because the interview yielded ample data bearing on mathematical and conceptual reasoning skills and his approaches to solving problems, enabling us to critically evaluate multiple explanations for his behavior when he got stuck.  We focused initially on the "getting stuck" episode because our project was focusing at that time on factors that impede mathematical sense-making. We should note that Jim was typical of our subjects in his ability to solve the problems we posed.

### Interview Protocol
To explore what facilitates or hinders engineering students' mathematical sense-making in introductory physics courses, the interviews focused on how students explain (to themselves and others) and use physics equations when solving physics problems.  In keeping with this objective, the interview context reflected that of the classroom in some ways:  Interviews occurred in the Physics building and addressed equations and concepts from their course.  In this way, we hoped that the obstacles to mathematical sense-making observed in the interview would relate to the obstacles cued by their course work, making our findings instructionally relevant.

Appendix A lists the complete interview protocol; but the conversational flow in particular moments suggested on-the-fly modifications to these prompts, or construction of new prompts, to probe the student's thinking more deeply.  In such moments, we prioritized pursuing the student's thoughts over strictly adhering to the interview protocol.  Of course, when analyzing our interview of Jim, the degree of standardization across different interviews becomes irrelevant.

The interview starts by asking students to explain an equation that they are familiar with: $v = v_0 + at$, referring to velocity, acceleration and time.  We tried to cue different pockets of students' knowledge with prompts such as, "How would you explain that equation to a friend from your English class?" or, "How would you explain that equation on a physics exam?" Subjects then solved a problem using that equation.  We followed up by asking the subjects to make sense of an equation that is similar in structure but new to the students at the time of the interview: the hydrostatic pressure equation, $p = p_{at\ top} + \rho gh$, where $p$ is the pressure under water in a lake or ocean, $p_{at\ top}$ is the pressure at the top, $\rho$ is the density of water, $g$ is the acceleration due to gravity and $h$ is the depth below the water's surface. We then posed a problem that could clearly be solved using that equation.

## Analysis

We began with close analysis of the raw video and transcript of the segment of interview where Jim gets stuck solving a problem using the hydrostatic pressure equation, and then, a few minutes later, when he gets unstuck.  Although we weren't building theory, our analysis borrowed from "grounded theory" [40],[41] the goal of generating possible explanations for why Jim got stuck and unstuck without relying on pre-conceived categories. However, we were inevitably influenced by our familiarity with possible explanation from previous work, including epistemological issues and skill deficits.  In trying to describe Jim's thinking, we attended not just to the literal meaning of his statements, but also to other markers of his thinking suggested by the discourse and framing analysis literature [42],[43], including word choice, pauses, pitch and register, body language, and so on.

After formulating plausible explanations of Jim's thinking in the selected snippets, we then searched the rest of the interview for segments that could help to confirm or disconfirm our explanations [44].

To increase the accuracy and accountability of our analyses, we did much of the above analysis in a larger (five-person) group, arguing about the snippets.  The five members commonly disagreed with each other.  Burkhardt and Schoenfeld [45] discuss the advantages of group analysis.  Second, we make the entire interview transcript available (NOTE TO EDITORS AND REVIEWERS: WE ARE TRYING TO ARRANGE FOR A SECURE PERMANENT SERVER FOR THIS) so that readers of this journal can check their own interpretations against ours.  Finally, in our arguments below, we present and analyze key segments of interview data, enabling the reader to evaluate our arguments against her own readings of the data.  For this reason, in what follows, we do not fully tease apart our data and analysis/discussion.

# Our focal episode:  Jim gets stuck

In this section, we present the focal episode of this article:  The interviewer, Ayush Gupta, asks whether the pressure 7 meters beneath the surface of a lake is greater than, less than, or equal to the pressure 5 meters beneath the surface.  Jim uses the equation $p = p_{at\ top} + \rho gh$, to which he had just been introduced, Jim makes a sign error and ends up concluding that the pressure is greater at a depth of 5 meters.  He sticks to this answer despite realizing that it violates the common-sense idea that the pressure is higher at greater depths.  In the following paragraphs, we present an overview of how Jim and Ayush interact trying to resolve the error. The transcript segments are time-stamped (min min: second second) .

**Jim makes a sign error**

When first posed the question, Jim spends a few seconds thinking quietly about it. Prompted by Ayush, he explains that he is trying to figure out how to plug the given quantities into the equation to compute the pressures at the two points. Ayush offers to let him know the value of any quantity, such as the density of water. At that moment, Jim expresses confusion about which sign to use for *h*:

> [24:01] Jim: Wait, it is under water, does under water mean that *h* is negative?
>
> Ayush: So it is set up so that *h* is the distance below the water.
>
> Jim: Which means it is negative.
>
> Ayush: Um,
>
> Jim: Depends on where your axis are
>
> Ayush: In this situation, the way *h* is mentioned, is that *h* is defined as going down from the water surface
>
> Jim: So then five is bigger and seven is smaller
>
> Ayush: Why do you say that?
>
> Jim: Because if it's going down then it is negative
>
> Ayush: What is negative?
>
> Jim: *h* is negative and this is a smaller number and five is a bigger number
>
> Ayush: Because of the negative?
>
> Jim: Yeah.
>
> [24:51]

Jim decides that *h* must be negative since it is increasing downward, possibly because of the convention often used in physics problems of taking the vertically down direction to be negative, and possibly due to miscommunication between him and Ayush.

**Jim sticks to his formal result and defends it**

Over the next six minutes, Jim sticks to the idea that *h* should be negative on account of its direction, and the pressure is greater at a depth of 5 m than at a depth of 7 m. For example, when the interviewer explicitly suggests that *h* be taken as positive, Jim agrees that it would lead to the conclusion that pressure is higher at a depth of 7m compared to 5m, but he seems uncomfortable with this idea and does not take it up.

[29:56] Ayush: Suppose I told you that h is positive.

Jim: Always positive?

Ayush: Yes, So I am taking the positive distance from the surface, but its considered positive. Would that help you?

Jim: I mean, that would just make 7 greater than 5.

Ayush: Okay. Does that bother you?

Jim: I mean... What I keep thinking is that you are going *down* (gestures down) so 7 cannot be greater than 5 and negative. That's why I keep coming back to that. Meaning, if you do say it's positive then ... I guess it doesn't bother me. (sighs) 7 is greater than 5 in positive-land.

[30:55]

Jim is skeptical about Ayush's suggestion that $h$ is positive, given that its direction is *down* from the surface. Jim sighs as he tries to accept that idea and perhaps tries to distance himself from it by saying it's true in "positive-land," a notion he introduces with sarcasm in his voice. In the conversation that follows, Jim reverts to the idea that $h$ should be negative.

Jim's "stickiness" in his conclusion that the pressure is greater at a shallower depth calls for explanation because he knows this answer contradicts common sense. Below, we present the short segment in which Jim describes how relying on everyday experience would lead to the conclusion that the pressure increases with depth. After that segment, Ayush probes whether Jim sees the contradiction:

[28:20] Ayush: So do you think that the mathematics here is telling you something different [from perceptual experience]?

Jim: Yeah, I think it is.

Ayush: Okay, okay. So suppose you were to answer this question on an exam, which one would you pick - the experience one or the math?

[28:35] Jim: I will pick the mathematics.

Below we present Jim's reason for choosing the mathematics. For now, our point is that, despite the interviewer's prompts, he never makes an attempt to reconcile the math with the common-sense answer, beyond double checking that he was using the hydrostatic pressure equation correctly.

In summary, Jim exhibits some common, previously-documented behaviors: not seeking a reconciliation between everyday experiences and formal mathematics — i.e., not mathematically sense-making — and favoring formal over everyday thinking [6],[24],[30]. Mathematical sense-making in this episode might have led Jim to take more seriously the possibility that $h$ is positive, since then the equation would agree with common sense, or to look for another way to reconcile common sense with his mathematical result.

We now turn to explaining why Jim does not successfully engage in mathematical sense-making during this episode.

## What does *not* explain Jim's behavior: Skill/knowledge deficits

In this section, we systematically rule out several plausible explanations for Jim's behavior, explanations based on his holding a robust misconception or on his lacking the needed knowledge or skills. Specifically, we will show that Jim's lack of mathematical sense-making in the above episode does *not* stem from:

(i)     misconceptions about coordinate systems (e.g., the idea that down must be negative),

(ii)    lack of (or incorrect) conceptions about pressure,

(iii)   underdeveloped mathematical manipulation skills,

(iv)    inability to connect math and physics ideas, or

(v)     unwillingness to revisit an answer, perhaps connected to unwillingness to admit he made a mistake.

The interview segments we presented above, and will present below, are not all in chronological order. To provide a sense of the overall conversational flow, and to situate the segments we analyze, we summarize the conversation in Table 1.

**(i) Jim *does not* have a fixed conceptualization of 'down as negative'**

Given Jim's persistent adherence to the idea that $h$ must be negative since "it is down from the surface of water," we might attribute to him an entrenched misconception about coordinate systems: that downward must always correspond to negative. Jim, however, does not possess this misconception. Evidence for this claim comes from other problem-solving episodes in the interview.

In one such episode, Jim is solving the "two-ball problem":

[07:16] Ayush: [H]ave a look at this problem here. um. it says that you are standing on the fourth floor and you are throwing the ball down with an initial speed of 2 m/s and you are letting go of another ball. okay. So, its asking you that after 5 seconds, what would be the difference in speeds of the two balls. Will it be less than, more than, or equal to 2 m/s?

In solving this problem, Jim productively, though tacitly, chose 'down' as the positive direction, assigning positive values to the initial velocity of the thrown ball and the acceleration due to gravity, *g*. This counts as evidence against his holding a misconception about coordinate systems, not simply because he takes downward as positive but because in doing so he avoids a coordinate-system error that students commonly make, treating the acceleration due to gravity as automatically negative, without checking for consistency with the other quantities.

Our data do not rule out the following more nuanced explanation of Jim's behavior: Certain cues cause Jim to activate a strong connection between his physical sense of "down-ness" and his mathematical sense of "minus." The two-ball problem does not activate this connection, perhaps because the salient feature of the problem to Jim is that the initial velocity and the acceleration caused by gravity go in the *same* direction; it's not salient to him in the moment that the direction happens to be downward. In the pressure scenario, by contrast, it is extremely salient to Jim that *h* is measured *downward* from the water's surface. By this account, Jim's difficulty arises not from a globally robust misconception about coordinate systems, or from a gap in his knowledge about coordinate systems, but rather, from the context-dependent strength of the connection between two elements of his thinking, his physical sense of downness and his mathematical sense of negative.

We actually endorse this account as a partial explanation of Jim's behavior. In doing so, we do not contradict our central claims, because:

(1) As just emphasized, we are not ascribing a knowledge deficit or a global misconception to Jim; in some contexts, he "knows" that downward can be positive, but he's not accessing that knowledge in the hydrostatic pressure problem.

(2) The coordinate-system-based explanation of Jim's behavior cannot be the whole story, for the following reason. As discussed below, Jim eventually *does* resolve the contradiction between his common sense and his mathematical result, in response to a prompt from Ayush. Jim achieves the reconciliation using reasoning tools and knowledge he already had available before Ayush's intervention, and without giving up the idea that *h* is negative. Epistemological considerations, we will argue, are what prevent Jim from finding that reconciliation on his own.

**(ii) Jim has sufficient conceptual understanding of pressure for this problem**
Based on Jim's defense of his result that greater depth corresponds to lower pressure, we might think that Jim lacks sufficient understanding of pressure to see the intuitive problem with his answer. But when asked how a friend from English class would reason, Jim shows an understanding that, intuitively speaking, pressure should increase with increasing depth:

[26:03] Ayush: Jim, suppose, um, let me ask another question. Suppose there was a friend of yours in English, right, not doing physics, so does not really know physics, and equations kind of thing. Could they have answered this question?
Jim: This question?
Ayush: Just he question that you know, under water, is the pressure greater than, less than, or equal to at a depth of 7 m versus a depth of 5m, could they answer that without really knowing physics?

Jim: Not unless they have experience being under water themselves. If they have, then they can.

Ayush: Okay. Then. So. What do you mean when you said that they have experience?

Jim: Like, if they have actually been under water, so the pressure, they might know a little bit about pressure under water.

Ayush: Umm ..

Jim: Like they have gone snorkeling under water

Ayush: What would they know?

Jim: Like a rough estimate. The pressure was higher when I was deeper. [Ayush: okay] The pressure was lower when I was higher to the surface. If they can actually work it in a equation, I do not think they will be able to.

Ayush: So, given that information, given that experience, could they have argued which pressure would be more - 7m or 5 m?

Jim: I mean.

Ayush: Not from equations maybe

Jim: Just from that. I mean they could argue it

Ayush: What would they argue then?

Jim: They could argue from their personal experience like, One time I was scuba diving and I was like 30 feet below the water and pressure was like, pressure was very high. Like I was just swimming, I was just couple of feet below the water and the pressure was not that much.

Ayush: So they would say that pressure at 7 m.

Jim: Yes, is greater. [Ayush: is greater.] But they are not factoring in the negative sign of h.   [28:17]

In brief, Jim's experiences indicate that the pressure at greater depths exceeds the pressure at shallower depths. We are claiming not that Jim has a deep understanding of pressure, but rather, that his understanding is more than adequate to answer the interview question correctly. He rejects his (correct) experience-based conclusion not because his experience-based reasoning is ambiguous or incorrect, but rather, because it is "not factoring in the negative sign of $h$."

**(iii) Jim fluidly performs mathematical manipulations**

Jim quickly and correctly reasoned that, assuming $h$ is positive (and all the constants are positive), the equation $p = p_{at\ top} + \rho g h$ implies that the pressure is greater at a depth of 7 meters than at a depth of 5 meters. He also solved the two-ball problem correctly, exhibiting no difficulties. Our point is that lack of mathematical problem-solving skills does not explain why Jim reaches and sticks with his incorrect conclusion about pressure.

**(iv) Jim can connect mathematical results to physical implications**

Another explanation for Jim's sticking with his conclusion that the pressure is greater at shallower depths could be that he has trouble relating mathematical conclusions to physical common sense. We saw above, however, that Jim knows that his mathematical reasoning, according to which greater depth corresponds to lower pressure, contradicts the intuitive answer that someone with "personal experience" under water would give.

**(v)  Jim is willing to change his mind**

Some of Jim's behavior while he's sticking to his mathematical answer (at the expense of the experience-based answer) suggests that he might be resisting the interviewer by "stubbornly" sticking to his initial answer.  Evidence supporting this explanation includes his sighing, sarcastic reaction to Ayush's suggestion that $h$ is positive ("7 is greater than 5 in positive-land"); his clenched posture during parts of the ensuing conversation; and his quick, forceful statement that he "will pick the mathematics" on an exam, when choosing between the mathematical and the experience-based answer, which contrasts with the long, thoughtful pauses that occur before many of his other answers throughout the interview. [NOTE TO EDITORS/REVIEWERS: WE WILL TRY TO POST VIDEO CLIPS FOR READERS; TRYING TO ARRANGE FOR SECURE SERVER].  If a conversational dynamic of opposition exists here, it could be enhanced by Jim's life experiences as an African American male in a field traditionally dominated by other racial/ethnic groups, in a K-12 and college schooling environment that in many cases ignores or even devalues cultural resources that African American students bring to bear [46].

We agree that the conversational dynamic in this segment of the interview, with Ayush continually prodding Jim to rethink his answer, probably contributes to Jim's "stickiness" in his answer.  But this stickiness cannot fully explain why he doesn't attempt to reconcile his mathematical and experience-based answer.  Our evidence is that Jim *does* change his mind, quickly and eagerly, in response to a particular prompt from Ayush:

> [34:40] Ayush: What do you think about $g$ in [the hydrostatic pressure] equation? Should that be minus ten or plus ten?
>
> Jim: Oh! minus ten ... So, that gives you a positive thing. [Ayush: Okay.] I would say that the negative does not matter anymore. Oooh! I see.  The higher you go under water, uh, the lower you go under water the more your pressure is, because the negative and the negative cancel out ... So, the more under water you are the higher your pressure is going to be, I think now. I forgot to factor in $g$. That's what I think.
>
> Ayush: Okay.  Is that more comfortable or less comfortable?
>
> Jim: That is more comfortable because it actually makes more sense to me now. And now your experience actually does work because from your experience being under water you felt more pressure as opposed to the surface. If I take into consideration both negatives, it is just positive, they just add up.  [36:08]

Notice that Jim not only gets excited about finding a reason to reverse his earlier conclusion about how pressure changes with depth ("Oh!", "Oooh!"), but also readily admits to making a mistake earlier:  "I forgot to factor in [the negative direction of] $g$."  Furthermore, he says that this revised answer "is more comfortable."  This is evidence that (i) Jim's earlier clenched posture stemmed at least in part from his feeling uncomfortable with his earlier answer, and (ii) any earlier apparent resistance he was exhibiting to rethinking his answer did not reflect a general unwillingness to revisit and revise his thinking or a robust stubbornness about changing his answer.

So, why *was* Jim so "sticky" in his conclusion that grater depth corresponds to lower pressure, at least until Ayush's prompt about the sign of *g*?  We take up this issue in the next section.

## What *does* explain Jim's behavior:  epistemology

In this section, we first argue that Jim gets stuck and stays stuck in his incorrect answer partly because he doesn't engage in key aspects of mathematical sense-making.  Then we argue that his lack of mathematical sense-making is driven in part by his epistemological stance.

### Jim doesn't engage in key aspects of mathematical sense-making

Mathematical sense-making — specifically, translating and reconciling between mathematical and everyday/common-sense reasoning — might have helped Jim (re)solve the pressure problem on his own.  Upon noticing the contradiction between his assumption that *h* is negative and the common-sense conclusion about how pressure depends on depth, he might have been more willing to rethink the conceptual meaning of *h* and hence to question his assumption about the sign of *h*, instead of brushing aside the idea that *h* could be positive.  This deeper questioning might have led him to tap into his knowledge, exhibited in the two-ball problem, that downward can correspond to positive.  Alternatively, thinking about why *g* appears in $p = p_{at\ top} + \rho gh$ might have led him to think about the sign of *g*.  Looking at the structure of the expression $p_{at\ top} + \rho gh$ and thinking about the physical implications of a negative $\rho gh$ — that the pressure everywhere under the water is *less* than the pressure at the top — might have led him to check his signs more carefully.  Of course, these instantiations of mathematical sense-making might not have worked for Jim.  Our point, however, is to explain why he didn't even *try* any of these avenues.  We argued in the last section that his failure to attempt deeper mathematical sense-making did not stem from misconceptions about coordinate systems, insufficient understanding of pressure, underdeveloped mathematical skills, inability to connect math and physics ideas, or unwillingness to change his answer.  Instead, we now argue, Jim's epistemology explains the absence from his reasoning of key features of mathematical sense-making.

### Jim's epistemological stance pushes him away from mathematical sense-making

During the interview, Jim (a) regards mathematical reasoning as much more trustworthy than everyday common-sense reasoning, and (b) does not see mathematical equations as expressing meaning that is tractably connected to everyday/common-sense reasoning.  These two features of his epistemological stance, we argue, help to cause and/or sustain a tendency to focus on the syntax of mathematical manipulations rather than the conceptual meaning of mathematical equations and expressions and how to reconcile that meaning with everyday common-sense reasoning.  In other words, Jim's epistemological stance, specifically (a) and (b) above, pushes him away from aspects of mathematical sense-making that might have helped him troubleshoot his reasoning without needing Ayush's prompt.

*(a) Jim regards mathematical reasoning as much more trustworthy than everyday reasoning*

Jim articulates the view that mathematical reasoning is much more trustworthy than everyday/common-sense reasoning immediately after describing how an English-major friend would respond to the question of whether the pressure is greater at 5 meters or 7 meters.  Recall

from above that, according to Jim, the English major would say the pressure is larger at greater depths:

[28:20] Ayush: So do you think that the mathematics here is telling you something different [from the English major's based-based answer]?

Jim: Yeah, I think it is.

Ayush: Okay, okay. So suppose you were to answer this question on an exam, which one would you pick - the experience one or the math?

Jim: I will pick the mathematics.

Ayush; Mathematics. Can you tell me why?

Jim: Uh, for mathematics. For an equation to be given to you it has to be like theory and it has to be fact-bearing. So, fact applies for everything. It is like a law. It applies to every single situation you could be in. But, like, your experience at times or perception is just different - or you don't have the knowledge of that course or anything. So, I will go with the people who have done the law and it has worked time after time after time. [29:15]

Here, Jim says that perceptions can be misleading at times, but that "for an equation to be given…it has to be fact-bearing" and hence "applies for everything…like a law." The reason to trust the equation as law-like is that "people…have done the law and it has worked time after time…" So, mathematics encode authoritatively verified truths that are much more trustworthy than a given person's perceptions.

Jim's statements and actions throughout the interview mesh with this epistemological view. Solving the two-ball problem, for instance, he considers his mathematical answer adequate without checking whether it agrees with common sense. And even with the pressure question, when Jim resolves the contradiction between mathematical formalism and the common-sense answer, the resolution is grounded in formalism: Jim concludes that since $g$ and $h$ are both negative, $\rho gh$ is positive and hence the pressure, $p = p_{at\ top} + \rho gh$, increases with depth. Nowhere in the interview does Jim express or enact the idea that common-sense reasoning could challenge or make him reconsider mathematical reasoning.

*(b) Jim does not see mathematical equations as expressing meaning that is tractably connected to everyday/common-sense reasoning*

As emphasized above, Jim sees his mathematical *answers* and his everyday/common-sense reasoning as talking to each other; they can agree or disagree, and if they disagree, a choice must be made. But he does not see mathematical equations as expressing meaning that tractably connects to his common-sense reasoning. In other words, he does not see mathematical equations in physics as expressing conceptual content that can inform or be informed by everyday/common-sense reasoning.

To be clear, we do not claim that Jim consciously holds this view. Epistemological stances can be tacit, exhibited through patterns of behavior [31],[47],[48]. Evidence for Jim's (tacit) epistemological stance comes mostly from the way he reasons with equations. The examples we now give reflect a pattern that was consistent throughout the interview.

When asked to explain the equation for the velocity of an object moving with constant acceleration, $v = v_0 + at$, Jim notes the equation can be obtained either by taking the derivative of a certain equation for position, or by integrating of a basic equation for acceleration:

[2:00] Jim: Velocity equals the initial velocity plus acceleration times time. The equation for speed and it comes from the derivative of position or the integral of acceleration. That is how I will explain it.

Several other interviewees, by contrast, started by defining the variables (as Jim did) but went on to try to explain why the equation makes sense, such as why acceleration is multiplied by time, or what the overall equation says in common sense terms (e.g., the final velocity is the initial speed plus however much velocity you gain because you're accelerating).

We can't attribute Jim's behavior here solely to the task structure of explaining rather than actually using an equation. When solving the two-ball problem using that equation, Jim adeptly performs mathematical manipulations to get the right answer: the thrown ball is moving 2 m/s faster than the dropped ball after they have both been falling for the same time. Ayush then asked if some one could have answered that question without all the calculations. Some other interviewees, in response to this follow-up prompt or in response to the original question, saw a shortcut stemming from the physical meaning of $v = v_0 + at$: since both balls *gain* the same velocity in a given time (with the gain given by $at$), and since the thrown ball starts out 2 m/s faster than the other (the difference in their $v_0$'s), the thrown ball is *still* going 2 m/s faster than the other 5 seconds later. But Jim said that without going through the math there was no way to get to an answer.

Similarly when asked how he would try to explain the hydrostatic pressure equation to himself, Jim checks that the dimensions work our properly — a productive first step in making sense of an equation. For Jim, however, it was the only step:

[13:05] Ayush: So this is just an equation about pressure; pressure inside water. a lake or ocean or something like that right. it just gives you an expression for that in terms of the pressure on top and then the density of the water and acc due to gravity, *g*, and the depth of water from the surface *h*. so suppose you were trying to understand this equation, you were trying to make sense of it. how would you go about doing that.

[For about six minutes: Jim works on the units of the temrs in the pressure equation; he asks Ayush about the units of pressure which Ayush provides him as Newtons per squared meters; he makes an error but recovers from it, and finally convinces himself that the terms in the equation have the same units.]

Jim: ah! it does match.

Ayush: are you happy?

Jim: Yeah. /smiles looking at his work/

Jim: What do they mean by how would you explain this equation to yourself? If I see this, I will first make sure that units match.

Ayush: Okay

Jim: What else do they mean by how would you explain this equation to yourself?

Ayush: Um, does this equation make sense to you?

Jim: The units make sense, so it has to be heading somewhere [okay]

Ayush: um So suppose we asked you to explain the equation on an exam?

Jim: work out the units and plug it in

Ayush: okay   [20:46]

So, while some other interviews tried to make sense of (for example) the $\rho g h$ term, Jim doesn't think about the conceptual meaning of that term, or of the overall structure of the equation, before concluding that the equation makes sense.

In this subsection, we've presented indirect evidence that Jim does not see mathematical questions as expressing conceptual meaning, or at least conceptual meaning that is tractably connected to everyday/common-sense reasoning.  The evidence has consisted of a consistent pattern in his use of equations. At one point in the interview, though, Jim comments on this issue more directly.  Taking on the perspective of the English major, he had just concluded that common-sense reasoning, based on part of experiences feeling squeezed under water, disagrees with his mathematically-based conclusion about the relation between depth and pressure.

[29:17] Ayush: So, okay, so, Jim, do you think this equation relates to [the physical experience of pressure]?

Jim: This one? (points to $p = p_{at\ top} + \rho g h$ on the page)

Ayush: Yes. Does it relate to that experience?

Jim: Probably somehow, but not directly.

Ayush: Can you tell me how?

Jim: I think there is some way that just completely links the two together, but it's not obvious what that relation is.   [29:50]

Here Jim acknowledges that there is "probably somehow…. some way that just links the two [mathematically-based conclusion and experiences under water] together," but the relation is not direct and "not obvious." This statement meshes with his consistent behavior in the interview of not looking for such connections.

*Jim's epistemological stance pushes him away from mathematical sense-making during the pressure episode*

We've just argued that Jim (a) regards mathematical reasoning as much more trustworthy than everyday common-sense reasoning, and (b) does not see mathematical equations as expressing meaning that is tractably connected to everyday/common-sense reasoning.  These two features of his epistemology reinforce each other in pushing him away from a key aspect of mathematical sense-making, systematically looking for meaning in and coherence between mathematical formalism and the system described by that formalism.  According to view (b), such meaning and coherence are unlikely to be found (at least by non-experts); and according to view (a), it's unnecessary to seek that kind of meaning and coherence, at least to solve problems, since the mathematical answer can be trusted at the expense of the common-sense answer.

We outlined above how mathematical sense-making might have helped Jim reach the correct answer to the pressure problem, and now we've offered a plausibility argument that Jim's epistemological stance causes and/or sustains his failure to engage in math sense-making in this episode.  More strikingly, we showed evidence of his epistemological stance affecting his behavior *in the moment* during that episode.  He concluded that $p = p_{at\ top} + \rho gh$ makes sense solely on the basis of dimensional analysis, cutting him off from also looking at the conceptual meaning of the equation.  He stuck with his mathematical answer even after realizing it starkly conflicts with common sense, citing the greater trustworthiness of math.  He didn't search harder for a resolution between his formal reasoning and common sense, and when asked why, cited the indirectness and non-obviousness of the connection between the two.  Of course, it could always be the case that his epistemological statements in this episode were rationalizations that didn't capture the epistemological stance (if any) actually driving his reasoning in those moments.  The plausibility of the causal connections we're inferring between his epistemological stance and his in-the-moment reasoning comes from the overall coherence and parsimony of our account:  the epistemological stance we inferred from the pressure episode also helps to account for Jim's reasoning during the rest of the interview, and the in-the-moment causal links we're inferring mesh not only with Jim's behavior but with previous case studies [6],[30] and larger-N correlational studies [25],[27]

## Revisiting Jim's epistemology:  nuance and seeds of expertise

As we've depicted it so far, Jim's epistemological stance is unproductive to his problem-solving and learning.  In classifying Jim's epistemology as unproductive or "unavailing" [8], we would be fitting into the general trend of literature on students' epistemologies, which typically assigns a globally-robust epistemological "level" to a student, at least within a given discipline [7]. Other researchers have argued, however, that even within a given discipline or within a given context (such as an interview), fine-grained contextual and conceptual cues can produce both subtle and dramatic shifts in a student's epistemological stance [33],[49],[50]. In this section we revisit Jim's epistemological stance to show that it incorporates more nuance, flexibility and seeds of expertise than we depicted above. We are not backing off our argument above that aspects of Jim's epistemological stance during the pressure episode impede his mathematical sense-making. We are arguing instead that even during this episode, Jim's epistemology exhibits productive seeds that well-designed instructional environments could help nurture towards epistemological expertise. As part of our argument, we'll show evidence that Jim underwent a subtle but important epistemological shift (as opposed to merely a conceptual shift) in response to Ayush's hint about the sign of $g$, a shift with important instructional implications. In this way, we avoid a deficit-based account of Jim's reasoning, an account that simply replaces deficient skills/knowledge deficiency with a "deficient" epistemology.

**Seeds of expertise in Jim's stance of trusting math as authoritative**

Above, we used the following quotation to argue that Jim views canonical mathematical equations as much more trustworthy than everyday/intuitive, on the grounds that such equations come from authoritative sources:

[28:39] Jim: For an equation to be given to you it has to be like theory and it has to be fact-bearing. So, fact applies for everything. It is like a law. It applies to every single situation you could be in. But, like, your experience at times or perception is just different - or you don't have the knowledge of that course or anything. So, I will go with the people who have done the law and it has worked time after time after time.

At first glance, Jim's stance here is one of accepting information from authority, specifically from "the people [presumably professional scientists or engineers] who have done the law and it has worked time after time." Accepting knowledge as absolute and coming from authority or direct personal observation is generally taken to be a low stage in someone's epistemological development [51],[52],[13]. But Jim's stance here is more nuanced. He is accepting the authority of those people *because* they have "done the law and *it has worked time after time.*" (emphasis added). So, the trustworthiness of the mathematical equation stems not from the credentials of people who promulgate it but rather from repeated empirical tests of the law. Like a professional scientist or engineer, Jim doesn't feel the need to replicate all these experiments for himself; repeated tests done by others is sufficient to make the equation trustworthy.

Nurturing these seeds towards expertise would involve helping Jim see himself as someone who can contribute to the formulation and testing of such equations; the epistemological resources he needs to understand the purpose and value of this kind of activity appear to already be in place.

**Seeds of expertise in Jim's views about the connection between formalism and common sense**

We showed above that, during the pressure episode before Ayush's *g* prompt, Jim doesn't expect mathematical equations to express meaning that is tractably connected to everyday/common-sense reasoning. Here, we show that this stance is nuanced and flexible; it contains seeds that, if further developed, will support mathematical sense-making. First, even before Ayush's prompt, Jim *does* expect a connection between mathematical formalism and common sense to exist, even if it's hard to track. Second, after Ayush's hint about the sign of *g*, Jim quickly shifts to a stance in which he expects common sense to mesh tractably with mathematical formalism, and he's clearly more comfortable with this stance. We now support these two points in turn.

*1. Jim expects a connection to exist between common sense and mathematical formalism.* Our evidence comes from the passage cited above in which Jim responds to Ayush's question about whether the pressure equation under consideration relates to the physical experience of pressure. While emphasizing the intractability of the connection, Jim nonetheless says the connection exists:

[29:35] Jim: I think there is some way that just completely links the two together, but it's not obvious what that relation is. (emphasis added).

This differs from a more naïve stance according to which mathematical reasoning and everyday reasoning exist in completely separate worlds.

*2. Jim shifts to a (subtly) different epistemological stance in response to a conceptual cue.*

As argued above, Jim was uncomfortable during the pressure episode, as indicated by long pauses, clenched body posture, and occasional annoyance or sarcasm. His comfort and excitement level changes dramatically, however, in response to the realization that he can reconcile the mathematical formalism with common sense:

[34:40] Ayush: What do you think about *g* in [the hydrostatic pressure] equation? Should that be minus ten or plus ten?

Jim: *Oh!* minus ten ... So, that gives you a positive thing. [Okay.] *I would say that the negative does not matter anymore. Oooh! I see.* The higher you go under water, uh, the lower you go under water the more your pressure is, because the negative and the negative cancel out ... So, the more under water you are the higher your pressure is going to be, I think now. I forgot to factor in *g*. That's what I think.

Ayush: Okay. Is that more comfortable or less comfortable?

Jim: *That is more comfortable because it actually makes more sense to me now. And now your experience actually does work because from your experience being under water you felt more pressure as opposed to the surface.* If I take into consideration both negatives, it is just positive, they just add up. (emphasis added).   [36:08]

Jim is excited that the problematic "negative [sign of *h*] does not matter any more," since the negative sign of *g* cancels it out.  He views this realization as an exciting insight — "Oh!", "Oooh! I see" — and gives us an indication of why:  "…because it actually makes sense to me now."  In this moment, he views his mathematical answer as making sense *because* it meshes with everyday experience, not simply because mathematical equations are trustworthy.

At first glance, this appears to be an affective shift but not an epistemological shift; after all, Jim had always maintained that a connection exists between mathematical formalism and everyday reasoning, and he is excited to have found that connection.  But a subtle epistemological shift has occurred as well.  Before Ayush's hint, the connection between mathematical formalism and common sense was an inert part of Jim's cognitive ecology; it didn't inform his reasoning, partly because he didn't expect that connection be something he could find.  After Ayush's hint, by contrast, the connection between mathematical formalism and common sense became an active element of Jim's cognition:  it makes him more comfortable than he was before, and drives his sense that his answer "actually makes more sense to me now."

A skeptic could argue that the shift we just described, with a formerly inert element of his epistemology suddenly becoming active, isn't really an epistemological *shift*: both before and after the so-called "shift," Jim held the same epistemological view, that everyday reasoning connects to mathematical formalism.  All that's changed, by this account, is his view of how tractable the connection is for the problem at hand, which reflects his confidence and knowledge, not his views about knowledge.  For the argument we're making, we need not contend with this description of Jim.  We are showing that Jim's epistemological stance contains productive seeds that, when developed, could contribute to his mathematical sense-making.  If Jim's view that a connection exists between formalism and everyday reasoning were always inert — if it never played an active role in his reasoning — then it could not serve as a productive seed.  Our point here is that Jim's view *does* play an active role in his reasoning, in some circumstances.  Whether or not the activation of this view counts as a "shift" is immaterial.

Instructionally nurturing this productive seed in Jim's epistemology could begin with helping him reconcile common sense with mathematical formalism over a range of situations, and encouraging him to reflect on his views about the connection between everyday and formal reasoning—with the goal of making this epistemological view active in Jim's thinking even when no reconciliation between formalism and everyday reasoning is readily apparent.

## Conclusion
We showed that Jim gets stuck and stays stuck on a hydrostatic pressure problem not because he lacks any of the relevant mathematical or conceptual skills/knowledge, but rather, because he does not engage in deep mathematical sense-making — specifically, translating and reconciling between mathematical and everyday/common-sense reasoning.  We argued that his epistemological stance helps to explain his lack of mathematical sense-making: Since he regards mathematical equations as much more trustworthy than everyday reasoning and does not view mathematical equations as expressing meaning that tractably connects to everyday reasoning, he does not view reconciling between common sense and mathematical formalism as either necessary or plausible to accomplish. However, Jim's epistemological stance contains productive seeds that instructors and instructional environments could nurture, to push him toward mathematical sense-making: He *does* see common-sense as connected to formalism and in some circumstances this connection is both salient and valued.

This case study has two main goals. First, it adds to previous studies about the link between epistemologies and mathematical problem-solving by showing how a student's epistemology affects his problem-solving in the moment: Jim's trust in formalism and his view that reconciling between common sense and formalism is intractably difficult contributes to his not only to failing to attempt such a reconciliation himself but immediately rejecting the idea from the interview (e.g., that h could be positive) that could have helped him achieve such a reconciliation. Second, it makes plausible the claim that helping Jim — and other students like him — take a more productive epistemological stance toward mathematical problem-solving need not involve helping him develop new epistemological beliefs from scratch. Rather, instructors can help students build on productive seeds already present in their epistemologies.

## Acknowledgements

We thank "Jim" for participating in the study. We thank the instructor for Jim's course for letting us recruit students from his class for interviewing. We thank David Hammer, Andrew Brantlinger, and Members of the Physics Education Research Group for discussions. This work was supported in part by NSF EEC-0835880 and REC-0440110.

## Author Biographies

Ayush Gupta is a Research Associate in the Department of Physics, University of Maryland, College Park. His research interests include complex systems models of students' reasoning and incorporating affect and identity in models of cognitive dynamics. Currently, Ayush is involved in two projects: (i) as the Project Director for NSF-funded project, "Improving Engineering Students' Mathematical Sense-making: Research and Development," and (ii) a professional development project aimed at helping elementary- and middle-school teachers engage their students in authentic scientific inquiry.

*Address:* Ayush Gupta, Room 1320, Toll Physics Building, University of Maryland, College Park, MD 20742.

Andrew Elby is an Assistant Research Scientist in the Department of Physics and the Department of Curriculum & Instruction, University of Maryland, College Park. Andy focuses on students' and teachers' epistemologies and their effect on learning and teaching. He currently leads an NSF-funded research and development project focused on engineering students' mathematical sense-making. In other research and development projects, he works with grade 4-8 teachers trying to engage their students in scientific inquiry and studies the intrapersonal variability in science students' and teachers stances toward knowledge.

*Address:* Andrew Elby, Toll Physics Building, University of Maryland, College Park, MD 20742.

## Tables
Table1: Timeline of Jim's interview.

| [00:06] | Conversation about study and consent form. |
|---|---|
| [01:46] | Jim explains the velocity equation to friend from English class. |
| [06:12] | Jim explains the velocity equation to a 12-year old. |
| [07:16] | Jim works on the two balls problems and gets the correct answer. |
| [11:55] | Jim answers if the problem can be solved without doing calculations. |
| [13:05] | Jim works on the units of the terms in the pressure equation. With some help, Jim finally convinces himself that the units do match. |
| [20:54] | Jim works quietly on comparing the pressures at depths of 5 m and 7m. |
| [24:03] | Jim asks if $h$ should be negative. Ayush's answer makes Jim think that "h" is negative and he concludes, "5 is greater, and 7 is smaller." |
| [24:51] | Jim sees that changing the sign of $h$ would change his conclusion, but when asked, he does not choose between the two options. |
| [26:01] | Jim explains that a friend from English class could argue that pressure increases with increasing depth under water. |
| [28:28] | Prompted, Jim says he would pick the mathematics result on an exam. Asked why, Jim explains that the mathematics results are more reliable. |
| [29:19] | Jim acknowledges there might be a connection between the equation and the physical experience of pressure. |
| [29:51] | Suggested that $h$ is positive, Jim makes the "positive-land" remark. |
| [30:55] | Jim says that a positive $h$ would make his friend happy "because then their theory makes sense." |
| [31:17] | When asked, Jim correctly infers physical implications of a negative $h$. |
| [32:38] | Asked to explain the equation to a 12-year old, Jim laughs and says that he would explain how perceptions might not always be right. |
| [34:33] | Alerted to the sign of "g" in the pressure equation, Jim immediately resolves the conflict around the pressure problem. |
| [36:14] | Jim correctly compares pressures at the same depth in water for a lake on Mars versus lake on Earth. |
| [37:29] | Brief, inconclusive conversation about comparing the pressures at the surfaces of the two lakes. |
| [39:22] | Jim explains that he would have to understand the units to really understand an equation, not just plug in numbers. |
| [41:22] | Jim says that algebra makes using mathematics difficult in physics course. |
| [43:50] | Jim works on the modified Atwood's machine problem. |
| [49:24] | Jim makes a correct intuitive argument about the acceleration of the hanging block. |
| [51:48] | Jim says, that photographic memory would help in grades but not in knowing the material, unless the photographic memory was eternal. |
| [53:12] | Jim says that that a person taking the course not for grades would still need to know the equations to understand the material. |
| [54:12] | Ayush asks what role equations play in physics. Jim expresses uncertainty about what that question means, but goes on to say that equations outline a step by step way to solve a problem. |

# Appendix A: Interview protocol

**Velocity Equation (V)**

(V1) Here's an equation you've probably seen in physics class: $v = v_0 + at$. How would you explain this equation to a friend from class?

(V2) How would you explain this on an exam? .... to a 12-year old?

**The Two Balls Problem (B)**

(B1) Suppose you are standing with two tennis balls in the balcony of a tall building. You throw one ball down with an initial speed of 2 m/s; You just let go of the other ball, i.e., just let it fall. I would like you to think aloud while figuring out what is the *difference in the speed of the two balls* after 5 seconds – is it less than, more than, or equal to 2 m/s? (Acceleration due to gravity is 10m/s$^2$)

(B2) Could you have answered this without doing the calculations?

**Hydrostatic Pressure Equation (P)**

(P1) Here's an equation you perhaps haven't yet learned.  It's a formula for the pressure at a given depth under the surface of a lake, ocean, or whatever: p = p$_{at\ top}$ + $\rho_{water}gh$, where p$_{at\ top}$ is the pressure at the surface of the water, $\rho_{water}$ is the density of water, and $h$ is the distance below the surface.  How would you explain that equation to yourself?

(P2) Is the pressure at h=5 meters under water greater than, less than or equal to the pressure at h=7 meters under water?

(P3) Consider a lake on the surface of Mars that has weaker gravity compared to earth. What that means is that "g" for Mars is lower than "g" for earth which is 10 m/s$^2$. Is the pressure at a depth in the earth-lake greater than, less than, or equal to the pressure at the same depth for the mars-lake?

**Newton's Law problem: Modified Atwood's machine (NL)**

(NL1) Here's a problem you may have encountered in physics. *(Frictionless everything.)*

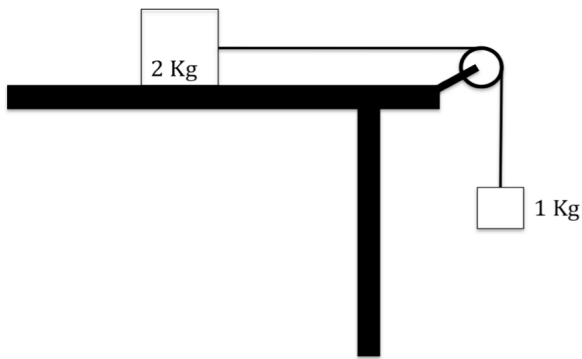

(NL2) Solve for the acceleration of the 1 kg block.  (*I would like you to think aloud as you are working on this. We're not interested in the answer you get but in how you think about it.* )

**Questions probing epistemology of equations in the physics context (E)**

(E1) How do you know when you really understand an equation?

(E2) What's hard about learning or using the math in this physics course?

(E3) Suppose you had photographic memory for equations.  Would that improve your performance?  Why?  *Follow-ups would try to tease apart whether the advantage is course specific or a more general failure of what it means to know math in a physics context.*

(E4) Suppose a student is taking the course for fun, and not getting graded, with the goal of understanding physics more deeply.  He's/She's not interested in learning to solve the quantitative problems, but he's willing to study outside of class to learn the concepts better.  What role if any should equations play in his studying?

(E5) Suppose you were given a list of equations on the exam – would that help you?